# Two-photon excitation of atoms by ultrashort electromagnetic pulses in a discrete spectrum

V.A. Astapenko, S.V. Sakhno

*Moscow Institute of Physics and Technology (State University)*

The development of methods of generation of ultrashort pulses (USP) of femto- and attosecond duration ranges with controlled parameters necessitates the theoretical study of features of their interaction with a matter [1]. Among such features that do not exist in case of "long" pulses should first of all be the nonlinear dependence of the photoprocess probability $W$ on the USP duration ($\tau$) [2] as well as the dependence on the carrier phase with respect to the pulse envelope ($\varphi$) [3-4]. It should be noted that if the dependence of the probability $W$ on the phase $\varphi$ manifests itself either only for very short pulses, when $\omega\tau < 1$ ($\omega$ is carrier frequency of the pulse), or in case of a nonlinear photoprocess [3], the function $W(\tau)$ can differ from a linear function in the limit $\omega\tau > 1$ too for fields of moderate strength, when the perturbation theory is applicable [2].

To describe photoprocesses in an USP field, various theoretical methods were used. Thus in the works of V.I. Matveev with co-authors [5-7] the sudden perturbation approximation was used to describe scattering of attosecond pulses by different quantum systems: atoms, ions, molecules, and clusters. In the V.P. Krainov's papers [8-9], excitation of a two-level system under the USP action was studied with the use of solution of the Schrödinger equation, and photoionization of atoms was calculated both within the framework of the perturbation theory [10] and in the Landau-Dykhne approximation [11]. In the latter work it was shown in particular that ionization of an atom by an intense single-cycle cosine pulse is much more efficient than under the action of a sine pulse.

In the paper [12], within the framework of the perturbation theory the formula was obtained that describes the total probability of single-photon absorption of an USP (during all time of its action) in terms of the spectral cross-section of photoabsorption and the Fourier transform of the strength of the electric field in a pulse. The expression derived in the work [12] was widely used further for analysis of single-photon absorption and spontaneous scattering by various targets [13-15].

The present work is dedicated to the theoretical analysis of two-photon excitation of atoms in a discrete energy spectrum by ultrashort electromagnetic pulses of femto- and subfemtosecond ranges of durations. As examples, excitation of hydrogen and sodium atoms from the ground state to excited states with a zero orbital moment is considered.

The amplitude of the two-photon transition $i \to f$ during the action of an electromagnetic pulse in the second order of the perturbation theory is given by the following expression:

$$A_{fi}^{(2ph)} = \left(-\frac{i}{\hbar}\right)^2 \int_{-\infty}^{\infty} dt' \int_{-\infty}^{t'} dt'' \langle f|\hat{V}(t')\hat{V}(t'')|i\rangle, \qquad (1)$$

where

$$\hat{V}(t) = -\hat{d}(t)E(t) \qquad (2)$$

is the operator of electromagnetic interaction in the form of a length, $\hat{d}(t)$ is the operator of the electric dipole moment of an atom in the interaction representation:

$$\hat{d}(t) = \exp(i\hat{H}_0 t/\hbar)\hat{d}\exp(-i\hat{H}_0 t/\hbar), \qquad (3)$$

$\hat{H}_0$ is the Hamiltonian of an unperturbed atom, $E(t)$ is the electric field strength. We assume an electromagnetic pulse to be linearly polarized and the usability condition for the dipole approximation to be fulfilled.

Using the expansion in terms of the complete system of functions $|n\rangle$, from the formulas (1) - (3) we find:

$$A_{fi}^{(2ph)} = \left(-\frac{i}{\hbar}\right)^2 \sum_n \int_{-\infty}^{\infty} dt' \int_{-\infty}^{t'} dt'' \exp(i\omega_{fn}t' + i\omega_{ni}t'') d_{fn} d_{ni} E(t')E(t''), \qquad (4)$$

where $d_{fn}, d_{ni}$ are the matrix elements of the electric dipole moment of an atom. Now let us express the electric field strengths in the right-hand side of the equation (4) in terms of their Fourier transforms $E(\omega')$ and $E(\omega'')$:

$$E(t') = \int_{-\infty}^{\infty} E(\omega')\exp(-i\omega' t')d\omega'/2\pi; \quad E(t'') = \int_{-\infty}^{\infty} E(\omega'')\exp(-i\omega'' t'')d\omega''/2\pi \qquad (5)$$

and make the substitution of the time variable:

$$t'' = t' + \tau, \qquad (6)$$

which will allow integration with respect to the time $t'$ with appearance of the delta function

$$\delta(\omega_{fi} - \omega' - \omega'') \qquad (7)$$

under the sign of integration with respect to the frequencies $\omega', \omega''$. In the formula (7), $\omega_{fi}$ is the transition eigenfrequency that is assumed to be a positive value.

The formula (7) describes the law of conservation of energy in excitation of the transition $i \to f$ by monochromatic components of the electric field of a pulse $E(\omega')$ and $E(\omega'')$:

$$\hbar \omega_{fi} = \hbar \omega' + \hbar \omega''. \qquad (8)$$

The delta function (7) and accordingly the equation (8) were obtained under the assumption that the spectrum of an electromagnetic pulse is considerably wider than the spectral width of the transition $i \to f$. This assumption is knowingly fulfilled for neutral atoms and femtosecond (and shorter) electromagnetic pulses.

It should be noted that if both frequencies $\omega', \omega''$ are positive, excitation of the transition occurs due to two-photon absorption, but if one of these frequencies is negative, there is stimulated Raman scattering. The negativeness of both frequencies is impossible since under the assumption $\omega_{fi} > 0$, and the equation (8) should be fulfilled. The presence of the delta function (7) allows integration with respect to the frequency $\omega''$, so under the integral with respect to $d\omega'$ the product of electric field strengths will remain: $E(\omega')E(\omega_{fi} - \omega')$. Assuming that the energies of intermediate states $|n\rangle$ have negative imaginary additives, it is possible to integrate with respect to the time variable $\tau$. As a result, for the probability of two-photon excitation of the bound-bound transition $i \to f$ during the action of a linearly polarized electromagnetic pulse ($\mathbf{E} // z$) we obtain the expression:

$$W_{fi}^{(2ph)} = \left|A_{fi}^{(2ph)}\right|^2 = \frac{1}{(2\pi\hbar)^2} \left|\int_{-\infty}^{\infty} d\omega' E(\omega')E(\omega_{fi}-\omega')M_{fi}^{(2)}(\omega')\right|^2, \qquad (9)$$

where

$$M_{fi}^{(2)}(\omega') = \frac{1}{\hbar} \sum_n \frac{d_{fn}^z d_{ni}^z}{\omega_{ni} - \omega' - i0} \qquad (10)$$

is the two-photon matrix element that for the transition between states with a zero orbital moment can be expressed in terms of the radial Green's function

$$M_{fi}^{(2)}(\omega') = \frac{2}{3} \langle f | r' g_{l=1}(r', r; \omega' - I_p/\hbar) r | i \rangle, \qquad (11)$$

where $I_p$ is the atomic ionization potential. In writing the equation (11), the selection rules are taken into account, from which it follows that in the case under consideration ($l_i = l_f = 0$) the contribution to the sum over intermediate states is made only by states with the quantum number of an orbital moment $l_n = 1$.

It should be noted that the expression (9) is meaningful in case of applicability of the perturbation theory, that is, when $W_{fi}^{(2ph)} < 1$.

In calculation of the probability of excitation of a hydrogen atom we will use the Sturmian expansion of the Coulomb Green's function that looks like (in at. u.):

$$g_{l=1}(r',r;E) = \frac{4rr'}{9\nu^3} \exp\left(-\frac{r+r'}{\nu}\right) \sum_{k=0}^{\infty} \frac{\Gamma(k+4) F(-k,4,2r/\nu) F(-k,4,2r'/\nu)}{k!(k+2-\nu)}, \quad (12)$$

where

$$\nu(E) = \frac{1}{\sqrt{-2E}}, \quad (13)$$

$F(a,b,x)$ is the confluent hypergeometric function, $\Gamma(z)$ is the gamma function.

In case of excitation of alkali metal atoms from the ground state, the main contribution to the two-photon matrix element is made by a discrete spectrum, and it will suffice to take into account the first excited $p$-state. Thus in excitation of a sodium atom at the transition $3s \to 4s$ it is possible to be limited to the contribution of the $4p$-state since the oscillator strength for the virtual transition $3s \to 4p$ is 0.99, that is, it practically "saturates" the sum rule.

In view of fine splitting of $np$-levels in alkali metal atoms, the approximate equation (in at. u.) for the two-photon matrix element can be obtained from the formula (10):

$$M_{fi}^{(2)}(\omega') \cong \sum_n \left\{ \frac{\sqrt{f_{fn}^{(1/2)}/|\omega_{fn}^{(1/2)}|} \sqrt{f_{ni}^{(1/2)}/\omega_{ni}^{(1/2)}}}{\omega_{ni}^{(1/2)} - \omega' - i0} + \frac{\sqrt{2 f_{fn}^{(3/2)}/|\omega_{fn}^{(3/2)}|} \sqrt{f_{ni}^{(3/2)}/\omega_{ni}^{(3/2)}}}{\omega_{ni}^{(3/2)} - \omega' - i0} \right\}, \quad (14)$$

where $f_{jn}^{(1/2),(3/2)}$, $\omega_{jn}^{(1/2),(3/2)}$ are the oscillator strengths and the eigenfrequencies of transitions. In the formula (14), the upper index ½ describes the states $nP_{1/2}^o$, and the index 3/2 describes the states $nP_{3/2}^o$. In derivation of (14), it was taken into account that the statistical weight of the initial and final states is 2 (with consideration for a spin), and the statistical weights of the states $nP_{1/2,3/2}^o$ are respectively 2 and 4.

Hereafter we will consider two-photon excitation of atoms under the action of a pulse of a corrected Gaussian shape (CGP). The CGP Fourier transform is [14]:

$$E_{cor}(\omega',\omega,\tau,\varphi) = iE_0 \tau \sqrt{\frac{\pi}{2}} \frac{\omega'^2 \tau^2}{1+\omega^2 \tau^2} \left\{ e^{-i\varphi-(\omega-\omega')^2 \tau^2/2} - e^{i\varphi-(\omega+\omega')^2 \tau^2/2} \right\}, \quad (15)$$

where $E_0$, $\tau$, $\omega$ are the amplitude, duration, and carrier frequency of a pulse, $\omega'$ is the "current" frequency, $\varphi$ is the carrier phase with respect to the envelope. From the formula (15) it follows in particular that the constant component of a CGP is equal to zero: $E_{cor}(\omega'=0)=0$.

For pulse durations $\tau \gg 1/\omega_{fi}$, the two-photon matrix element $M_{fi}^{(2)}(\omega')$ can be factored outside the integral sign in the formula (9) at the frequency $\omega' = \omega_{fi}/2$ with retention of accuracy for the function $W_{fi}^{(2ph)}(\tau)$ within several percents:

$$W_{fi}^{(2ph)}(\tau \gg 1/\omega_{fi}) \cong \frac{1}{\hbar^2}\left|M_{fi}^{(2)}(\omega_{fi}/2)\int_{-\infty}^{\infty}d\omega' E(\omega')E(\omega_{fi}-\omega')\right|^2. \quad (16)$$

After calculation of the integral in the right-hand side of the equation (16) we obtain:

$$W_{fi}^{(2ph)}(\tau \gg 1/\omega_{fi}) \cong \frac{\pi}{2^{12}}\left|M_{fi}^{(2)}(\omega_{fi}/2)\frac{E_0^2}{\hbar}\tau\right|^2 \frac{(\omega_{fi}^4\tau^4 - 4\omega_{fi}^2\tau^2 + 12)^2}{(1+\omega^2\tau^2)^4}\exp\left[-(\omega-\omega_{fi}/2)^2\tau^2\right]. \quad (17)$$

From the formula (17) it is possible to determine the pulse duration corresponding to the maximum of the excitation probability at specified values of frequencies $\omega$ and $\omega_{fi}$ (in at. u.):

$$\tau_{max}(\omega) = \frac{f_m(r=\omega_{fi}/\omega)}{\omega}, \quad (18)$$

where

$$f_m(r) \cong \frac{1}{\sqrt{2}|r-2|}\sqrt{2-(r-2)^2 + \sqrt{(2-(r-2)^2)^2 + 40(r-2)^2}}. \quad (19)$$

For $r = 2$ the pulse duration at the maximum becomes infinite, which corresponds to the monotonically increasing dependence of the probability $W_{fi}^{(2ph)}(\tau)$ at $2\omega = \omega_{fi}$.

Let us introduce the normalized probability of two-photon excitation:

$$\widetilde{W}_{fi}^{(2ph)} = \frac{W_{fi}^{(2ph)}}{E_0^4}, \quad (20)$$

where the amplitude of the electric field strength $E_0$ is measured in atomic units.

The results of calculations by the above formulas of the normalized probability of two-photon excitation of the transition $1s \to 2s$ in a hydrogen atom under the action of a CGP are given in Figs. 1, 2. The similar results for a sodium atom for the transition $3s \to 4s$ are presented in Figs. 3, 4.

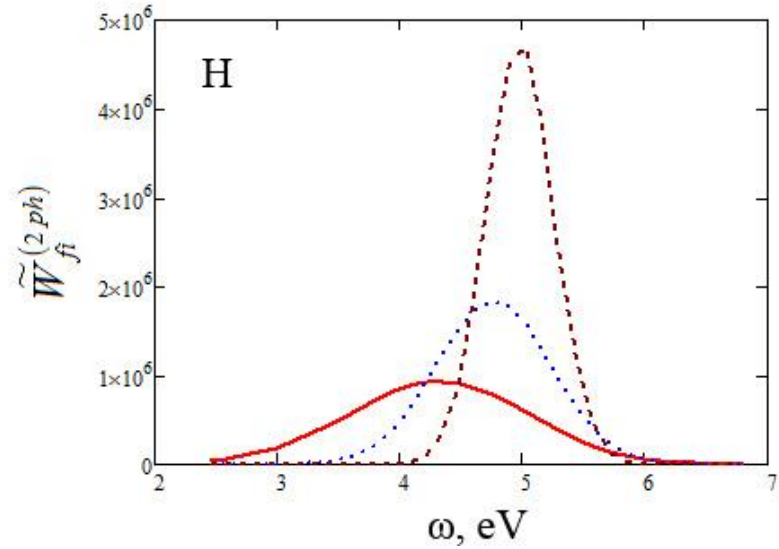

Fig. 1. The spectrum of the normalized probability of two-photon excitation of a hydrogen atom at the transition 1s-2s for different CGP durations: solid curve - $\tau = 0.48$ fs, dotted curve - $\tau = 0.72$ fs, dashed curve - $\tau = 1.2$ fs

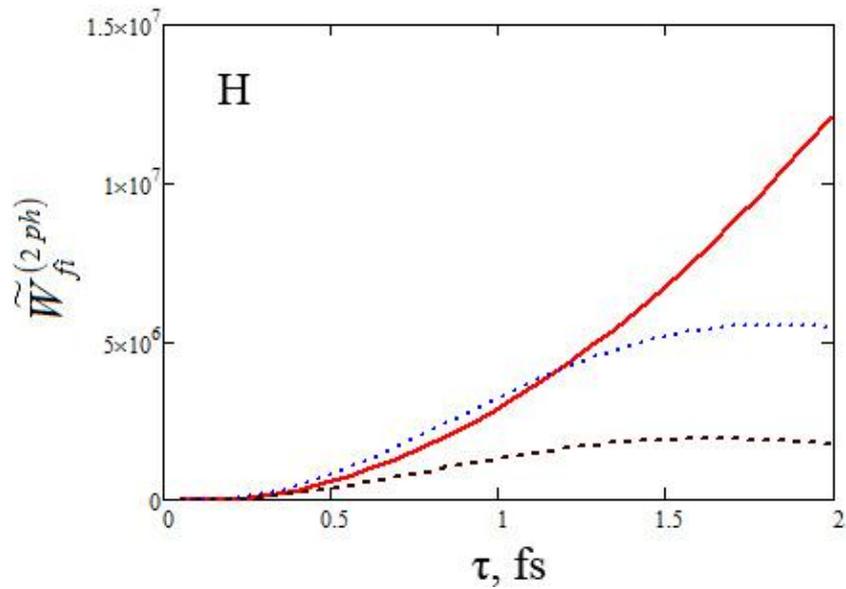

Fig. 2. The normalized probability of two-photon excitation of a hydrogen atom at the transition 1s-2s as a function of pulse duration for different carrier frequencies of a CGP: solid curve – $\omega = 5.1$ eV (two-photon resonance), dotted curve - $\omega = 4.84$ eV, dashed curve - $\omega = 5.39$ eV

It should be noted that for $E_0 = 10^{-2}$ at. u. the representative absolute value of the probability of two-photon excitation of a hydrogen atom for the problem parameters presented in Figs. 1-2 is $10^{-5}$.

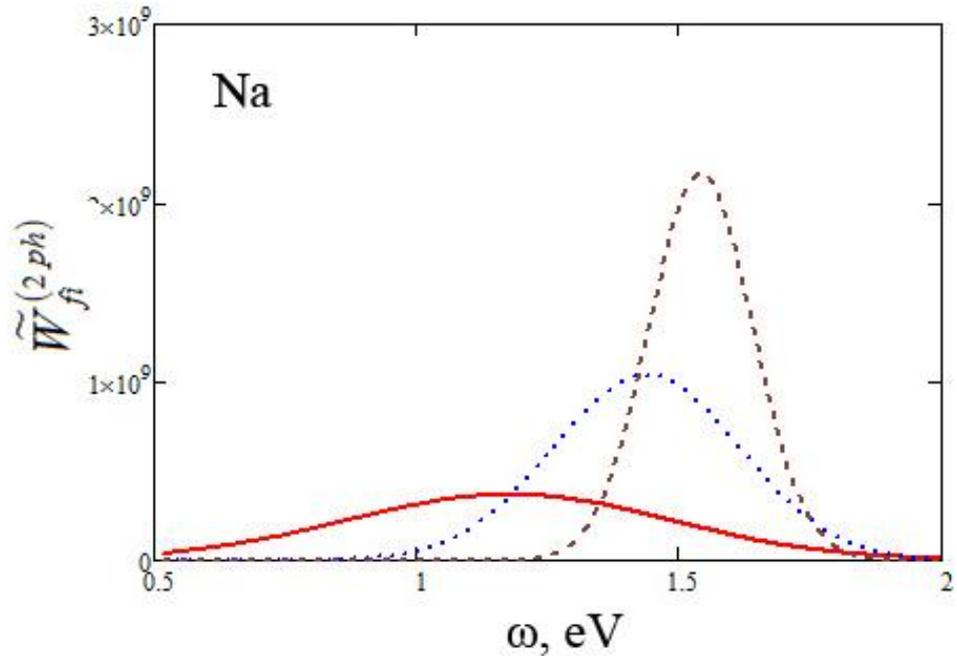

Fig. 3. The spectrum of the normalized probability of two-photon excitation of a sodium atom at the transition 3s-4s for different CGP durations: solid curve - $\tau = 1.2$ fs, dotted curve - $\tau = 1.92$ fs, dashed curve - $\tau = 3.36$ fs

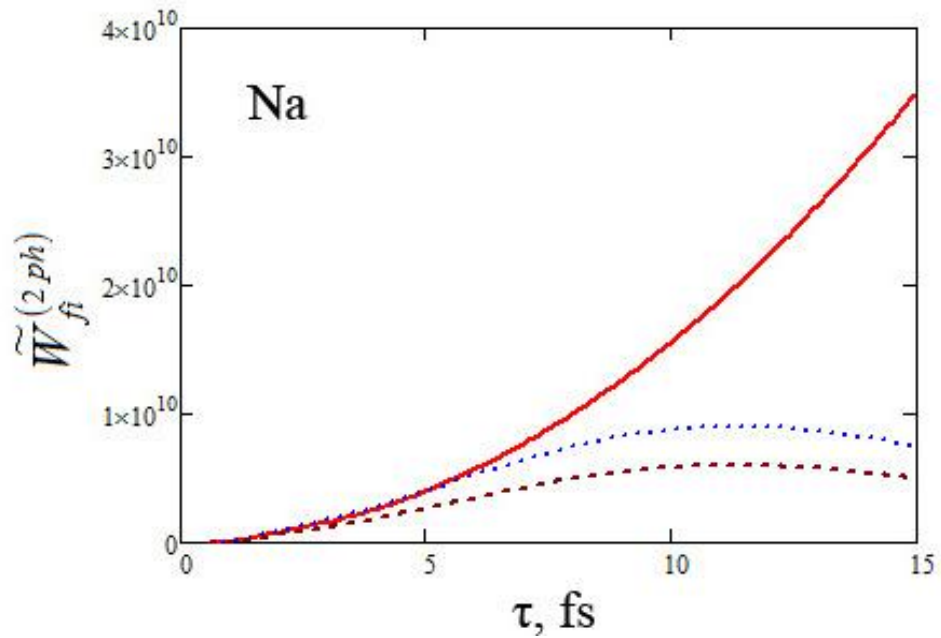

Fig. 4. The normalized probability of two-photon excitation of a sodium atom at the transition 3s-4s as a function of pulse duration for different carrier frequencies of a CGP: solid curve - $\omega = 1.59$ eV (two-photon resonance), dotted curve - $\omega = 1.55$ eV, dashed curve - $\omega = 1.63$ eV

From Figs. 1-4 it follows that the spectral and time dependences of the probability of two-photon excitation of atoms under consideration by femto- and subfemtosecond pulses are similar and differ only by numerical values (in case of sodium, the probability is three orders of magnitude more due to the presence of an approximate resonance at the virtual transition 3s-4p). Thus in both cases the spectrum of the excitation probability is broadened with decreasing pulse duration, and the spectral maximum in this case is shifted to the region of lower carrier frequencies. The dependence of the probability of two-photon excitation on the CGP duration (for $2\omega \neq \omega_{fi}$) is a curve with a maximum, the position of which is shifted to the region of long durations when the carrier frequency approaches the half transition frequency. In case of fulfilment of the two-photon resonance condition $2\omega = \omega_{fi}$, the excitation probability monotonically increases with pulse duration. The dependence of the pulse duration at the maximum on the carrier frequency for the transition 3s-4s in a sodium atom calculated by the formulas 18-19 is presented in Fig. 5.

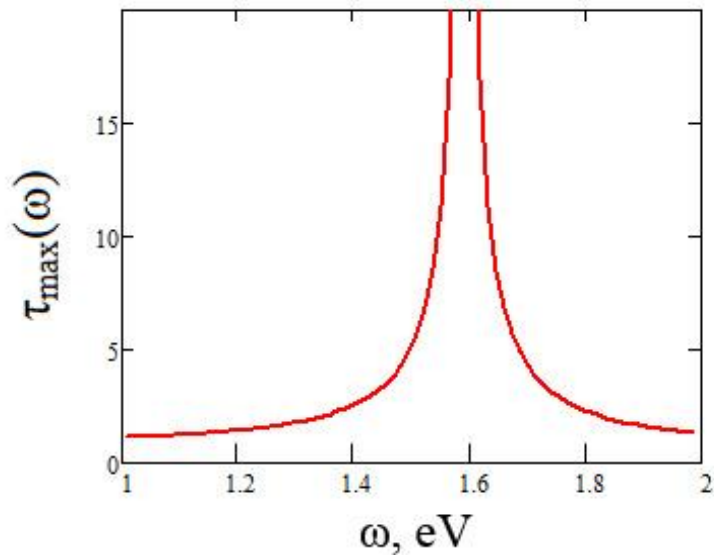

Fig. 5. The CGP duration at the maximum of the probability of two-photon excitation of the transition 3s-4s in a sodium atom as a function of carrier frequency, the ordinate is plotted in femtoseconds, the abscissa is plotted in electron-volts

From Fig. 5 it follows that when the carrier frequency approaches the resonance frequency, the pulse duration at the maximum of the probability of two-photon excitation increases indefinitely.

In the present work, the features of two-photon excitation of atoms in a discrete spectrum under the action of ultrashort electromagnetic pulses were studied theoretically. An expression for the probability of two-photon excitation of a bound-bound transition during the action of a linearly polarized electromagnetic pulse was obtained. Based on this expression, a case of excitation of hydrogen and sodium atoms was considered. The results of calculations have shown that corresponding spectral and time characteristics for these atoms are similar. A decrease in pulse duration results in spectral broadening and shift of the maximum of the spectral dependence to the region of lower values of carrier frequencies, and a decrease in the peak value of the probability at the maximum. In a nonresonance case ($2\omega \neq \omega_{fi}$), the probability of two-photon excitation as a function of pulse duration is a curve with a maximum, the position of which, with decreasing carrier frequency, is shifted to the region of long times and is increased in amplitude. In case of a two-photon resonance ($2\omega = \omega_{fi}$), the dependence goes to a monotonically increasing function.